\documentclass[twocolumn,showpacs,preprintnumbers,amsmath,amssymb]{revtex4}

\usepackage{graphicx}
\usepackage{dcolumn}
\usepackage{bm}

\newcommand{\bq}{\begin{equation}}
\newcommand{\eq}{\end{equation}}
\newcommand{\bqa}{\begin{eqnarray}}
\newcommand{\eqa}{\end{eqnarray}}
\newcommand{\nn}{\nonumber \\}

\def\be     {\begin{equation}}
\def\ee     {\end{equation}}
\def\bea        {\begin{eqnarray}}
\def\eea        {\end{eqnarray}}
\def\bnn    {\begin{eqnarray*}}
\def\enn    {\end{eqnarray*}}

\begin{document}

\title{Deconfined quantum criticality in the two dimensional Kondo lattice model}
\author{Ki-Seok Kim}
\affiliation{ School of Physics, Korea Institute for Advanced
Study, Seoul 130-012, Korea }
\date{\today}

\begin{abstract}
We investigate the continuous quantum phase transition from an
antiferromagnetic metal to a heavy fermion liquid based on the
Kondo lattice model in two dimensions. We propose that
antiferromagnetic spin fluctuations and conduction electrons
fractionalize into neutral bosonic spinons and charged spinless
fermions at the quantum critical point. This deconfined quantum
criticality leads us to establish a critical field theory in terms
of the fractionalized fields interacting via emergent U(1) gauge
fields. The critical field theory not only predicts non-Fermi
liquid physics near the quantum critical point but also recovers
Fermi liquid physics away from the quantum critical point.
\end{abstract}


\maketitle

Nature of quantum criticality is one of the central interests in
modern condensed matter physics. Especially, {\it deconfined
quantum criticality} has been proposed in various strongly
correlated electron systems such as low dimensional quantum
antiferromagnetism\cite{Senthil_deconfinement,Kim1,
Ichinose_deconfinement,Review,Wen_QED3,Hermele_QED3,
Kleinert_QED3,Kim2,Kim3} and heavy fermion
liquids\cite{Senthil_Kondo,Coleman_Kondo,Pepin_Kondo,Kim_Kondo}.
In the present paper we focus our attention on the quantum phase
transition from an antiferromagnetic metal to a heavy fermion
liquid in the two dimensional Kondo lattice model. In the
Landau-Ginzburg-Wilson theoretical frame work the quantum phase
transition would belong to the first order because the two phases
are characterized by two different order
parameters\cite{Senthil_deconfinement}. However, it is well known
that there exists a continuous quantum phase transition between
the two
phases\cite{Senthil_Kondo,Coleman_Kondo,Pepin_Kondo,Kim_Kondo}. In
this paper we propose that {\it the continuous quantum phase
transition can be realized via deconfined quantum criticality,
where antiferromagnetic spin fluctuations and conduction electrons
fractionalize into neutral bosonic spinons and charged spinless
fermions at the quantum critical point}. Near the quantum critical
point two kinds of critical fluctuations are expected to arise.
One would correspond to critical fluctuations of Kondo singlets
and the other, critical antiferromagnetic spin fluctuations.
Remarkably, {\it the emergent spinless fermions are associated
with critical fluctuations of the Kondo singlets while the
critical bosonic spinons result from critical spin fluctuations}.

We consider the Kondo lattice model in two dimensions \bqa && H =
H_{c} + H_{m} + H_{K} , \nn && H_{c} =
-t\sum_{<i,j>}c_{i\sigma}^{\dagger}e^{iA_{ij}}c_{j\sigma} - h.c. ,
\nn && H_{m} = J\sum_{<i,j>}{\vec S}_{i}\cdot{\vec S}_{j} , \nn &&
H_{K} = J_{K}\sum_{i}{\vec S}_{i}\cdot{c}_{i\sigma}^{\dagger}{\vec
\tau}_{\sigma\sigma'}c_{i\sigma'} . \eqa $H_c$ describes dynamics
of conduction electrons $c_{i\sigma}$ and $H_m$, that of local
spins ${\vec S}_{i}$. Here $t$ is a hopping integral of the
conduction electrons and $J$, an antiferromagnetic exchange
coupling of the local spins. $A_{ij}$ is an external
electromagnetic vector potential in a lattice version. $H_K$
represents Kondo couplings between the conduction and localized
spins, where $J_{K}$ is a Kondo coupling constant.

We perform the Haldane mapping for $H_m$ to derive the O(3)
nonlinear $\sigma$ model as a low energy effective field
theory\cite{Auerbach,Nagaosa}. In this mapping high energy
antiferromagnetic spin fluctuations would induce new interactions
between low energy spin degrees of freedom and conduction
electrons. We can derive the following expression for the Kondo
lattice model   \bqa && Z = \int{Dc_{i\sigma}}{D{\vec
n}_{i}}\delta(|{\vec n}_{i}|^{2} - 1)e^{- S_{m} - S_{K} - S_{c}} ,
\nn && S_{m} =
iS\sum_{i}(-1)^{i}\int_{0}^{c\beta}{dx_{0}}\int_{0}^{1}{du}{\vec
n}_{{i}}\cdot\Bigl(\frac{\partial{{\vec
n}_{{i}}}}{\partial{u}}\times\frac{\partial{{\vec
n}_{{i}}}}{\partial{x_0}}\Bigr) \nn && + \int_{0}^{c\beta}{dx_{0}}
\Bigl[ \frac{1}{2g}\sum_{i}|\partial_{0}{\vec n}_{i}|^{2} -
\frac{1}{g}\sum_{<i,j>}\vec{n}_{i}\cdot\vec{n}_{j} \Bigr] , \nn &&
S_{K} = - i\frac{2a^{d}J_K}{cg} \sum_{i}\int_{0}^{c\beta}{dx_{0}}
\Bigl(\frac{\partial{\vec n}_{i}}{\partial{x}_{0}}\times{\vec
n_{i}}\Bigr)\cdot{c}^{\dagger}_{i\sigma}{\vec
\tau}_{\sigma\sigma'}{c}_{i\sigma'}   \nn && +
\frac{2\sqrt{2d}a^{d-1}J_{K}}{cg}\sum_{i}\int_{0}^{c\beta}{dx_{0}}(-1)^{i}{\vec
n}_{i}\cdot{c}_{i\sigma}^{\dagger}{\vec
\tau}_{\sigma\sigma'}c_{i\sigma'} , \nn && S_{c} =
\int_{0}^{c\beta}{dx_0} \Bigl[
\sum_{i}c^{\dagger}_{i\sigma}(\partial_{0} - iA_{i0} -
\mu)c_{i\sigma} \nn && -
t_{c}\sum_{<i,j>}c_{i\sigma}^{\dagger}e^{iA_{ij}}c_{j\sigma} -
h.c. -
\frac{a^{2d}J_{K}^{2}}{c^{2}g}\sum_{i}|c^{\dagger}_{i\sigma}{\vec
\tau}_{\sigma\sigma'}c_{i\sigma'}|^{2} \Bigr] . \nn \eqa The local
spins are described by the O(3) nonlinear $\sigma$ model of a low
energy spin variable ${\vec n}$, where $g =
\frac{2\sqrt{2d}}{S}a^{d-1}$ is a spin stiffness parameter
($g^{-1}$) and $c = \sqrt{2d}JS{a}$, a velocity of spin
waves\cite{Auerbach,Nagaosa}. $a$ is a lattice spacing, $S$, the
value of localized spins, and $d$, a spatial dimension. Here $S =
1/2$ and $d = 2$. The first term in $S_m$ is a Berry phase action,
where $u$ and $x_0$ are two parameters in a unit
sphere\cite{Auerbach,Nagaosa}. $x_{0} = c\tau$ is considered to be
a rescaled time. The mapping to the nonlinear $\sigma$ model
results in nontrivial couplings between the low energy spins and
the conduction electrons. The first term in $S_K$ arises from high
energy spin fluctuations while the second originates from the
usual Kondo couplings. Notice $(-1)^{i}$ in the second term. In
the action $S_{c}$ for the conduction electrons $\mu$ is a
chemical potential of the conduction electrons and $A_{0i}$, an
external Coulomb potential. The hopping integral is redefined to
be $t_{c} = t/c$. The last term in $S_c$ represents local
ferromagnetic interactions for the conduction electrons,
originating from the contribution of high energy spin fluctuations
in the Haldane mapping.

The next step would be to integrate over high energy conduction
electrons. It is not easy to integrate over the conduction
electrons. Considering our objective to establish a critical field
theory {\it in terms of the fractionalized fields}, we are
satisfied with estimating possible effects of the Kondo couplings.
{\it The second Kondo coupling term in $S_K$ would reduce the spin
stiffness $g^{-1}$}. In the absence of the Kondo couplings the
nonlinear $\sigma$ model exhibits an antiferromagnetic long range
order\cite{Auerbach,Nagaosa}. It is well known that the O(3)
nonlinear $\sigma$ model shows a continuous quantum phase
transition from antiferromagnetism to paramagnetism at a nonzero
critical spin stiffness, $g_{c}^{-1}$ in two space and one time
dimensions\cite{Auerbach,Nagaosa}. Thus, increasing the Kondo
coupling constant $J_K$ may drive the spin stiffness $g^{-1}$ to
the critical value $g^{-1}_{c}$, causing the magnetic quantum
phase transition. {\it We view the continuous quantum phase
transition in the Kondo lattice model as the magnetic one induced
by the Kondo interactions}. Furthermore, the first Kondo coupling
term in $S_K$ would affect the Berry phase action in $S_m$. It is
well known that in two leg ladders the contribution of Berry phase
cancels between the legs\cite{Ladder,Kim_1D}. The same mechanism
works in the double layered quantum
antiferromagnets\cite{Kim1,Chubukov}. {\it The presence of the
Kondo couplings with the conduction electrons are expected to
weaken the contribution of Berry phase}.

Recently, it was argued that {\it at the quantum critical point of
the O(3) nonlinear $\sigma$ model, spin $1$ critical
antiferromagnetic fluctuations break up into more elementary spin
$1/2$ critical bosonic excitations called
spinons}\cite{Senthil_deconfinement,Kim1, Ichinose_deconfinement}.
This is the precise meaning of deconfined quantum criticality in
the quantum antiferromagnetism. {\it In the present paper we apply
the deconfined quantum criticality in the magnetic quantum phase
transition to the present Kondo lattice model because the quantum
phase transition in the Kondo lattice model is conjectured to be a
magnetic transition}. This idea can be realized in the $CP^{1}$
representation ${\vec n} = \frac{1}{2}z^{\dagger}_{\sigma}{\vec
\tau}_{\sigma\sigma'}z_{\sigma'}$, where $z_{\sigma}$ is a bosonic
spinon\cite{Senthil_deconfinement,Kim1,Ichinose_deconfinement,Auerbach,Nagaosa}.
A problem is the fate of conduction electrons near the quantum
critical point. In order to investigate this we focus on the
second Kondo coupling term ${\vec
n}_{i}\cdot{c}_{i\sigma}^{\dagger}{\vec
\tau}_{\sigma\sigma'}c_{i\sigma'}$ in $S_{K}$. Using ${\vec
n}\cdot{\vec \tau} = U\tau^{3}U^{\dagger}$ with an SU(2) matrix
$U = \left( \begin{array}{cc} z_{\uparrow} & -z_{\downarrow}^{\dagger} \\
z_{\downarrow} & z_{\uparrow}^{\dagger} \end{array} \right)$, we
can represent the Kondo interaction term as
${c}_{i\sigma}^{\dagger}
(U_{i}\tau^{3}U_{i}^{\dagger})_{\sigma\sigma'}c_{i\sigma'}$. This
term can be solved by introducing the following gauge
transformation $\chi_{\sigma} =
U^{\dagger}_{\sigma\sigma'}c_{\sigma'}$. The components of the
$\chi_{\sigma}$ field are given by $\chi_{\sigma} =
\left(\begin{array}{c} \chi_{\uparrow}
\\ \chi_{\downarrow} \end{array} \right) = \left(\begin{array}{c}
z^{\dagger}_{\uparrow}c_{\uparrow}
+  z_{\downarrow}^{\dagger}c_{\downarrow} \\
-z_{\downarrow}c_{\uparrow} + z_{\uparrow}c_{\downarrow}
\end{array} \right)$, where the $\chi_{\uparrow}$ field represents the
usual Kondo hybridization and the $\chi_{\downarrow}$ field
denotes the singlet pairing between the bosonic spinon and the
conduction electron. {\it The $\chi_{\sigma}$ field can be
considered to express Kondo resonances}. Near the quantum critical
point the bosonic spinons $U^{\dagger}_{\sigma\sigma'}$ emerging
from localized spins are expected to {\it hybridize} with the
conduction electrons $c_{\sigma'}$ via {\it the Kondo
interactions}, resulting in charged spinless fermions
$\chi_{\sigma}$. Another way to say this is that the conduction
electrons $c_{\sigma}$ fractionalize into the bosonic spinons
$U_{\sigma\sigma'}$ and the charged spinless fermions
$\chi_{\sigma}$, i.e., $c_{\sigma} =
U_{\sigma\sigma'}\chi_{\sigma'}$ owing to the Kondo interactions.

Representing Eq. (2) in terms of the fractionalized fields,
$U_{\sigma\sigma'}$ and $\chi_{\sigma}$, we can obtain the
following expression \bqa && Z =
\int{D\chi_{\sigma}}{DU_{\sigma\sigma'}}e^{-S_{c}-S_{m}} , \nn &&
S_{c} = \int {d^{3}x} \Bigl[
\chi^{\dagger}_{\sigma}\Bigl([\partial_{0} -
iA_{0}]\delta_{\sigma\sigma'} +
[U^{\dagger}\partial_{0}U]_{\sigma\sigma'} \nn && + {\vec
v}_{F}\cdot([i{\vec \nabla} + {\vec k}_{F} + {\vec
A}]\delta_{\sigma\sigma'} + i[U^{\dagger}{\vec
\nabla}U]_{\sigma\sigma'}) \Bigr)\chi_{\sigma'} \Bigr] \nn && +
\frac{2\sqrt{2d}a^{d-1}J_{K}}{c\tilde{g}}
\int_{0}^{c\beta}{dx_{0}} \sum_{k}
{\chi}_{k\sigma}^{\dagger}\tau^{3}_{\sigma\sigma'}\chi_{k+Q\sigma'}
, \nn && S_{m} = \int{d^3x} \Bigl[
Tr\Bigl(\frac{1}{4\tilde{g}}|\partial_{\mu}(U\tau^3U^{\dagger})|^{2}\Bigr)
\Bigr]  . \eqa Notice that the bare spin stiffness $g^{-1}$ is
assumed to renormalize into $\tilde{g}^{-1}$, resulting from the
contribution of high energy conduction electrons via the Kondo
couplings. In the action $S_{c}$ for the conduction electrons we
performed the continuum limit near the Fermi surface. Here $v_{F}$
is a Fermi velocity and $k_{F}$, a Fermi wave vector. The main
point in $S_{c}$ is that the effect of couplings between
antiferromagnetic spin fluctuations and conduction electrons
appears in the kinetic energy of the $\chi_{\sigma}$ field via the
gauge transformation $\chi_{\sigma} =
U^{\dagger}_{\sigma\sigma'}c_{\sigma'}$. The last term in $S_{c}$
originating from the Kondo couplings now represents the
contribution of Zeeman energy under staggered magnetic fields,
where $Q$ is a momentum $(\pi,\pi)$. The staggered Zeeman term
splits the two Kondo resonance states. In the limit of large
Zeeman splitting the $\chi_{\uparrow}$ fields describe the Kondo
resonances at odd sites while the $\chi_{\downarrow}$ fields do at
even sites. In the O(3) nonlinear $\sigma$ model we ignored the
Berry phase action because the first Kondo coupling term of $S_K$
in Eq. (2) is expected to weaken the contribution of Berry phase
as the case of the double layered quantum
antiferromagnets\cite{Kim1,Ladder,Kim_1D,Chubukov}. The first
Kondo coupling term of $S_K$ in Eq. (2) is also neglected. This
term does not affect critical phenomena because the bare scaling
dimension of $\Bigl(\frac{\partial{\vec
n}_{i}}{\partial{x}_{0}}\times{\vec
n_{i}}\Bigr)\cdot{c}^{\dagger}_{i\sigma}{\vec
\tau}_{\sigma\sigma'}{c}_{i\sigma'}$ is larger than $3$. Thus, the
scaling dimension of the coupling constant $\frac{a^{d}J_K}{cg}$
is negative, indicating that the couplings vanish in the low
energy limit. The local ferromagnetic interactions of $S_{c}$ in
Eq. (2) can be also ignored in the low energy limit. In the spinon
representation this term is $-
\frac{a^{2d}J_{K}^{2}}{c^{2}g}|c^{\dagger}_{\sigma}{\vec
\tau}_{\sigma\sigma'}c_{\sigma'}|^{2} = -
\frac{a^{2d}J_{K}^{2}}{c^{2}\tilde{g}}|\chi^{\dagger}_{\sigma}(U^{\dagger}{\vec
\tau}_{\sigma\sigma'}U)_{\sigma\sigma'}\chi_{\sigma'}|^{2}$. Since
the scaling dimension of
$|\chi^{\dagger}_{\sigma}(U^{\dagger}{\vec
\tau}_{\sigma\sigma'}U)_{\sigma\sigma'}\chi_{\sigma'}|^{2}$ is
much larger than $3$, this term vanishes in the low energy limit.

{\it The effective action Eq. (3) has not only the trivial
electromagnetic $U_{A}(1)$ gauge symmetry but also a new emergent
${U}_{a}(1)$ gauge symmetry}. This $U_{a}(1)$ gauge symmetry
guarantees the invariance of the action Eq. (3) under the gauge
transformations of \bqa && \chi'_{\sigma} =
[e^{i\vartheta\tau^{3}}]_{\sigma\sigma'}\chi_{\sigma'} \mbox{,
}\mbox{ }\mbox{ }\mbox{ }\mbox{ } U'_{\sigma\sigma'} =
U_{\sigma\alpha}[e^{-i\vartheta\tau^{3}}]_{\alpha\sigma'} , \nn &&
c'_{\sigma} = c_{\sigma}  \mbox{, }\mbox{   }\mbox{   }\mbox{
}\mbox{   } A'_{\mu} = A_{\mu} . \eqa This local gauge symmetry
implies that there should be a new emergent U(1) gauge field
corresponding to the gauge symmetry. Indeed, performing some
standard algebra such as the Hubbard-Stratonovich
transformation\cite{Kim_AF_ASL}, we can see the emergence of new
U(1) gauge fields $a_{\mu}$ \bqa && Z_{U} =
\int{DU_{\sigma\sigma'}}exp\Bigl[ - \int{d^3x} \Bigl(
Tr(\frac{1}{4\tilde{g}}|\partial_{\mu}(U\tau^3U^{\dagger})|^{2})
\nn && + \chi^{\dagger}_{\sigma}(
[U^{\dagger}\partial_{0}U]_{\sigma\sigma'} + i{\vec
v}_{F}\cdot[U^{\dagger}{\vec \nabla}U]_{\sigma\sigma'})
\chi_{\sigma'} \Bigr) \Bigr] \nn && =
\int{DU_{\sigma\sigma'}}{Da_{\mu}}exp\Bigl[ - \int{d^3x} \Bigl(
Tr(\frac{1}{2\tilde{g}}|(\partial_{\mu}-
ia_{\mu}\tau^{3})U^{\dagger}|^{2}) \nn && -
ia_{0}\chi^{\dagger}_{\sigma}\tau^{3}_{\sigma\sigma'}\chi_{\sigma'}
+ {\vec a}\cdot{\vec
v}_{F}\chi^{\dagger}_{\sigma}\tau^{3}_{\sigma\sigma'}
\chi_{\sigma'} \nn && -
\frac{\tilde{g}}{2}|\chi^{\dagger}_{\sigma}\tau^{3}_{\sigma\sigma'}\chi_{\sigma'}|^{2}
+ \frac{\tilde{g}}{2}|{\vec
v}_{F}\chi^{\dagger}_{\sigma}\tau^{3}_{\sigma\sigma'}\chi_{\sigma'}|^{2}
\Bigr) \Bigr] . \eqa Combining Eq. (3) and Eq. (5), we reach an
effective field theory for the quantum phase transition in the
Kondo lattice model \bqa && Z =
\int{D\chi_{\sigma}}{DU_{\sigma\sigma'}}{Da_{\mu}}e^{-S_{c}-S_{m}}
, \nn && S_{c} = \int {d^{3}x}\Bigl[
\chi^{\dagger}_{\sigma}\Bigl([\partial_{0} -
iA_{0}]\delta_{\sigma\sigma'} - ia_{0}\tau^{3}_{\sigma\sigma'} \nn
&& + {\vec v}_{F}\cdot([i{\vec \nabla} + {\vec k}_{F} + {\vec
A}]\delta_{\sigma\sigma'} + {\vec a}\tau^{3}_{\sigma\sigma'})
\Bigr)\chi_{\sigma'} \nn && -
\frac{\tilde{g}}{2}|\chi^{\dagger}_{\sigma}\tau^{3}_{\sigma\sigma'}\chi_{\sigma'}|^{2}
+ \frac{\tilde{g}}{2}|{\vec
v}_{F}\chi^{\dagger}_{\sigma}\tau^{3}_{\sigma\sigma'}\chi_{\sigma'}|^{2}
\Bigr] \nn && + \frac{2\sqrt{2d}a^{d-1}J_{K}}{c\tilde{g}}
\int_{0}^{c\beta}{dx_{0}} \sum_{k}
{\chi}_{k\sigma}^{\dagger}\tau^{3}_{\sigma\sigma'}\chi_{k+Q\sigma'}
, \nn && S_{m} = \int{d^3x} \Bigl[
Tr\Bigl(\frac{1}{2\tilde{g}}|(\partial_{\mu} -
ia_{\mu}\tau^{3})U^{\dagger}|^{2}\Bigr)  \Bigr] . \eqa The compact
U(1) gauge field $a_{\mu}$ guarantees the $U_{a}(1)$ local gauge
symmetry, where the gauge field is transformed into $a'_{\mu} =
a_{\mu} + \partial_{\mu}\vartheta$ under the transformations in
Eq. (4). From Eq. (6) the bosonic spinons
$U^{\dagger}_{\sigma\sigma'}$ and the spinless fermions
$\chi_{\sigma}$ can be considered to carry an internal charge
$e_{a}$ associated with the internal U(1) gauge field $a_{\mu}$.
Especially, the spinless fermions have not only the internal
charge $e_{a}$ but also the real electric charge $e_{A}$. The
kinetic energy of the SU(2) matrix $U^{\dagger}$ can be expressed
into the familiar $CP^{1}$ representation,
$({1}/{2\tilde{g}})Tr\Bigl(|(\partial_{\mu} -
ia_{\mu}\tau^{3})U^{\dagger}|^{2}\Bigr) =
({1}/{\tilde{g}})|(\partial_{\mu} - ia_{\mu})z_{\sigma}|^{2}$.

We would like to emphasize that {\it Eq. (6) is just another
representation of Eq. (3) derived from Eq. (2) via the gauge
transformation $\chi_{\sigma} =
U^{\dagger}_{\sigma\sigma'}c_{\sigma'}$}. For this rewriting to be
{\it physically} meaningful beyond a mathematical derivation,
there should exist the deconfined quantum criticality of the
nonlinear $\sigma$ model allowing deconfined bosonic spinons at
the quantum critical point. In other words, at least the nonlinear
$\sigma$ model in the effective field theory Eq. (6) should be
stable {\it in the renormalization group sense}. Fortunately, this
deconfined quantum criticality was shown to exist in Refs.
\cite{Senthil_deconfinement,Kim1} by a renormalization group
analysis, as mentioned before. Based on the exact transformation
and the existence of the deconfined quantum criticality, we
discuss the quantum critical point between an antiferromagnetic
metal and a heavy fermion liquid from Eq. (6). When the
deconfinement of bosonic spinons is not allowed owing to instanton
effects of compact U(1) gauge fields
$a_{\mu}$\cite{Senthil_deconfinement,Kim1}, the effective field
theory Eq. (6) becomes unstable in the present decoupling scheme
and thus, another effective theory necessarily results. This
indeed happens away from the quantum critical point, i.e., in an
antiferromagnetic metal and a heavy fermion liquid, as will be
discussed.

The gauge transformation introduced for solving the Kondo coupling
term in Eq. (1) may be still suspected to be unnatural although
similar decoupling schemes have been utilized in Refs.
\cite{Coleman_Kondo,Pepin_Kondo}. In this respect it is necessary
to understand the present methodology more deeply by comparing
this with other well studied ones. A good example is a $d-wave$
BCS theory for superconductivity of high $T_c$
cuprates\cite{Fisher_Z2,Tesanovic}. In the context of $d-wave$
superconductivity the coupling term of
$|\Delta|e^{i\phi}c_{\uparrow}c_{\downarrow}$ between Cooper pairs
and electrons plays the same role as the Kondo coupling term of
${\vec n}\cdot{c}_{\sigma}^{\dagger}{\vec
\tau}_{\sigma\sigma'}c_{\sigma'}$ between spin fluctuations and
conduction electrons. Here $|\Delta|$ and $\phi$ are the amplitude
and phase of Cooper pair fields. In order to solve this coupling
term several kinds of gauge transformations are
introduced\cite{Fisher_Z2,Tesanovic}. In these decoupling schemes
critical phase fluctuations of Cooper pairs screen out charge
degrees of freedom of electrons, causing electrically neutral but
spinful electrons called "spinons". As a result the phase factor
disappears in the coupling term when it is rewritten in terms of
spinons. Instead, this coupling effect appears as current-current
interactions of neutral spinons and phase fields of Cooper pairs
in the kinetic energy of electrons. Depending on the gauge
transformations, $Z_2$\cite{Fisher_Z2} or U(1)\cite{Tesanovic}
gauge fields are obtained. {\it In this respect the present gauge
transformation naturally extends the methodology of charge U(1)
symmetry in the context of superconductivity to that of spin SU(2)
symmetry in the context of antiferromagnetism}.

As mentioned earlier, the O(3) nonlinear $\sigma$ model $S_{m}$ in
Eq. (6) exhibits the continuous quantum phase transition at the
nonzero critical spin stiffness
$g_{c}^{-1}$\cite{Auerbach,Nagaosa}. Remember that the spin
stiffness can be controlled by the Kondo coupling constant
$J_{K}$. We define the critical Kondo coupling $J_{K}^{c}$ leading
to the critical spin stiffness $g_{c}$. In the case of $\tilde{g}
< g_{c}$ ($J_{K} < J_{K}^{c}$) the condensation of bosonic spinons
occurs, $<U^{\dagger}_{\sigma\sigma'}> \not= 0$ ($<z_{\sigma}>
\not= 0$), resulting in antiferromagnetism. The spinon
condensation leads the gauge field $a_{\mu}$ to be massive
(Anderson-Higgs mechanism). In the context of gauge theories this
phase corresponds to the Higgs-confinement
phase\cite{Fradkin,NaLee}. {\it The condensed bosonic spinons
$U_{\sigma\sigma'}$ are confined with the spinless fermions
$\chi_{\sigma'}$ to make the original conduction electrons
$c_{\sigma}$, i.e., $U_{\sigma\sigma'}\chi_{\sigma'} \rightarrow
c_{\sigma}$}. Performing the unitary gauge $a_{\mu}\tau^{3} =
a'_{\mu}\tau^{3} - iU\partial_{\mu}U^{\dagger}$ and the gauge
transformation $\chi_{\sigma} =
U^{\dagger}_{\sigma\sigma'}c_{\sigma'}$ in Eq. (6), and
integrating over the massive gauge field $a'_{\mu}$, we obtain the
following field theory for the conduction electrons in the
presence of antiferromagnetism (${\vec n}\cdot{\vec \tau} =
\tau^{3}$), \bqa && S_{c} = \int{d^3x}\Bigl[
c^{\dagger}_{\sigma}\Bigl( [\partial_{0} - iA_{0}] + {\vec
v}_{F}\cdot[i{\vec \nabla} + {\vec k}_{F} + {\vec A}]
\Bigr)c_{\sigma} \Bigr] \nn && +
\frac{2\sqrt{2d}a^{d-1}J_{K}}{c\tilde{g}}
\int_{0}^{c\beta}{dx_0}\sum_{k}
{c}_{k\sigma}^{\dagger}\tau^{3}_{\sigma\sigma'}c_{k+Q\sigma'} .
\eqa Deep inside the antiferromagnetism, the staggered Zeeman term
would vanish owing to $J_{K} \rightarrow 0$ in the renormalization
group sense. Thus, usual Fermi liquid physics is recovered. This
phase is an antiferromagnetic metal, where the antiferromagnetism
and Fermi liquid physics are nearly separated except the staggered
Zeeman term.

\begin{table*}
\caption{Quantum phase transition in the Kondo lattice model}
\begin{tabular}{cccccccc}
\hline & Antiferromagnetic Metal & Quantum Critical Point & Heavy
Fermion Liquid \nn & $\tilde{g} < g_{c}$ & $\tilde{g} = g_{c}$ &
$\tilde{g} > g_{c}$ \nn \hline
  Order Parameter & $<U^{\dagger}_{\sigma\sigma'}> \not= 0$
  ($<z_{\sigma}> \not= 0$) & & $<U^{\dagger}_{\sigma\sigma'}> = 0$ ($<z_{\sigma}> = 0$) \nn
  Internal Charge ($e_a$) & $U_{\sigma\sigma'}\chi_{\sigma'} \rightarrow c_{\sigma}$
  & $c_{\sigma} \rightarrow U_{\sigma\sigma'}\chi_{\sigma'}$
  & $c_{\sigma} \rightarrow U_{\sigma\sigma'}\chi_{\sigma'}$ \nn
  & $z^{\dagger}_{\sigma}\tau^{\pm}_{\sigma\sigma'}z_{\sigma'} \rightarrow \pi^{\pm}$
  & & $z^{\dagger}_{\sigma}{\vec \tau}_{\sigma\sigma'}z_{\sigma'} \rightarrow {\vec n}$
  \nn & Confinement of $e_a$ & Deconfinement of $e_a$ & Confinement of $e_a$
  \nn Gapless Excitations & $\pi^{\pm}$, $c_{\sigma}$  & $U^{\dagger}_{\sigma\sigma'}$ ($z_{\sigma}$),
  $a_{\mu}$, $\chi_{\sigma}$ & $\chi_{\sigma}$   \nn
  \hline
\end{tabular}
\end{table*}

Approaching the quantum critical point $\tilde{g} \rightarrow g_c$
($J_{K} \rightarrow J_{K}^{c}$), critical antiferromagnetic
fluctuations of the low energy localized spin variable ${\vec n}$
would fractionalize into critical bosonic spinons $z_{\sigma}$
($U^{\dagger}_{\sigma\sigma'}$)\cite{Senthil_deconfinement,Kim1,
Ichinose_deconfinement}. {\it An important point is that owing to
the nonzero critical coupling the bosonic spinons are expected to
screen out the spin degrees of freedom of the conduction electrons
$c_{\sigma}$, making the spinless fermions (Kondo resonances)
$\chi_{\sigma} = U^{\dagger}_{\sigma\sigma'}c_{\sigma'}$}. The
internal charge $e_a$ would be deconfined owing to critical
fluctuations of the bosonic
spinons\cite{Senthil_deconfinement,Kim1,Ichinose_deconfinement}.
Furthermore, dissipative dynamics of the gauge field arising from
the contribution of non-relativistic fermions $\chi_{\sigma}$ in
the presence of the Fermi
surface\cite{Senthil_Kondo,Kim_Kondo,Tsvelik_gauge,Nagaosa_gauge,Ioffe_gauge,C_v1,C_v2}
would increase the tendency of
deconfinement\cite{Kim3,Nagaosa_damping}. Thus, both the critical
spinless fermions and bosonic spinons would appear at the quantum
critical point. The resulting critical field theory is obtained to
be from Eq. (6) \bqa && S_{QCP} = \int{d^3x}\Bigl[
\chi^{\dagger}_{\sigma}\Bigl([\partial_{0} -
iA_{0}]\delta_{\sigma\sigma'} - ia_{0}\tau^{3}_{\sigma\sigma'} \nn
&& + {\vec v}_{F}\cdot([i{\vec \nabla} + {\vec k}_{F} + {\vec
A}]\delta_{\sigma\sigma'} + {\vec a}\tau^{3}_{\sigma\sigma'})
\Bigr)\chi_{\sigma'} \Bigr] \nn && +
\frac{2\sqrt{2d}a^{d-1}J_{K}}{c\tilde{g}}
\int_{0}^{c\beta}{dx_{0}} \sum_{k}
{\chi}_{k\sigma}^{\dagger}\tau^{3}_{\sigma\sigma'}\chi_{k+Q\sigma'}
\nn && + \int{d^3x} \Bigl[
Tr\Bigl(\frac{1}{2\tilde{g}}|(\partial_{\mu} -
ia_{\mu}\tau^{3})U^{\dagger}|^{2}\Bigr) \Bigr]  . \eqa  In the
above we ignored the local interactions since they are irrelevant
at the quantum critical point. This critical field theory is
expected to show non-Fermi liquid physics owing to long range
gauge
interactions\cite{Senthil_Kondo,Kim_Kondo,Tsvelik_gauge,Nagaosa_gauge,Ioffe_gauge,C_v1,C_v2}.
Critical fluctuations of the bosonic spinons result in the
non-Maxwell kinetic energy of the gauge field,
$\frac{N_z}{16}(\partial\times{a})\frac{1}{\sqrt{-\partial^2}}(\partial\times{a})$,
where $N_z$ is the flavor number of the bosonic spinons, here
$N_{z} = 2$\cite{Kleinert}. This leads to the following effective
action \bqa && S_{QCP} = \int{d^3x}\Bigl[
\chi^{\dagger}_{\sigma}\Bigl([\partial_{0} -
iA_{0}]\delta_{\sigma\sigma'} - ia_{0}\tau^{3}_{\sigma\sigma'} \nn
&& + {\vec v}_{F}\cdot([i{\vec \nabla} + {\vec k}_{F} + {\vec
A}]\delta_{\sigma\sigma'} + {\vec a}\tau^{3}_{\sigma\sigma'})
\Bigr)\chi_{\sigma'} \nn &&
+\frac{N_z}{16}(\partial\times{a})\frac{1}{\sqrt{-\partial^2}}(\partial\times{a})
\Bigr] \nn && + \frac{2\sqrt{2d}a^{d-1}J_{K}}{c\tilde{g}}
\int_{0}^{c\beta}{dx_{0}} \sum_{k}
{\chi}_{k\sigma}^{\dagger}\tau^{3}_{\sigma\sigma'}\chi_{k+Q\sigma'}
. \eqa In the absence of the last Zeeman term this effective
action was shown to give $C_{v} \sim TlnT$ and $\sigma \sim
T^{-5/3}$ in two dimensions, where $C_{v}$ and $\sigma$ are
specific heat and conductivity, respectively, and $T$,
temperature\cite{Kim_Kondo}. Remember that in the case of the
Maxwell kinetic energy the specific heat is given by $C_{v} \sim
T^{2/3}$ in two dimensions\cite{Senthil_Kondo,Tsvelik_gauge} and
$C_{v} \sim TlnT$ in three
dimensions\cite{Senthil_Kondo,C_v1,C_v2}. The conductivity is
shown to be $\sigma \sim T^{-4/3}$ in two dimensions and $\sigma
\sim T^{-5/3}$ in three
dimensions\cite{Senthil_Kondo,Nagaosa_gauge,Ioffe_gauge} in the
case of the Maxwell kinetic energy. We can see that the unusual
dynamics of the gauge field described by the non-Maxwell kinetic
energy results in "three dimensional effect"\cite{Kim_Kondo}. The
non-Fermi liquid physics is not captured in the antiferromagnetic
metal, where gauge fluctuations are suppressed via the
Anderson-Higgs mechanism. The presence of the staggered Zeeman
term would not give rise to qualitative changes in the non-Fermi
liquid physics. This term is expected to make the $\chi_{\sigma}$
band flatter, renormalizing the Fermi velocity $v_{F}$ and the
Fermi momentum $k_{F}$. A little heavier $\chi_{\sigma}$ fields
would arise from the Zeeman term. $C_{v} \sim TlnT$ and $\sigma
\sim T^{-5/3}$ are expected to remain at the deconfined quantum
critical point.

In the case of $\tilde{g} > g_{c}$ ($J_{K} > J_{K}^{c}$) the
bosonic spinons are gapped, $<U^{\dagger}_{\sigma\sigma'}> = 0$
($<z_{\sigma}> = 0$). Deep inside the quantum disordered
paramagnetism, the spin stiffness $g^{-1}$ would vanish in the
renormalization group sense\cite{Auerbach}. The vanishing spin
stiffness gives rise to one problem that the strength of local
interactions in Eq. (6) goes to infinity. Infinitely strong
interactions are expected to suppress fluctuations of the
$\chi_{\sigma}$ fermions. In order to treat the infinitely strong
local interactions, we perform the Hubbard-Stratonovich
transformation to obtain
$i\alpha_{0}\chi^{\dagger}_{\sigma}\tau^{3}_{\sigma\sigma'}\chi_{\sigma'}
+ i{\vec \alpha}\cdot{\vec
v}_{F}\chi^{\dagger}_{\sigma}\tau^{3}_{\sigma\sigma'}\chi_{\sigma'}$,
where $\alpha_{\mu}$ is an auxiliary field. Integration over the
$\alpha_{\mu}$ field leads to the local constraint of
$\chi^{\dagger}_{\sigma}\tau^{3}_{\sigma\sigma'}\chi_{\sigma'} =
0$. This local constraint makes the gauge field $a_{\mu}$ decouple
to the spinless fermions. Another way to say this is that the
$\chi_{\sigma}$ fields do not screen out the internal charge $e_a$
owing to the local constraint. Only the Maxwell kinetic energy is
available, resulting in confinement of gapped bosonic spinons.
Gapped paramagnons are expected to arise from the confinement. The
resulting low energy action is given by \bqa && S_{c} =
\int{d^3x}\Bigl[ \chi^{\dagger}_{\sigma}\Bigl( [\partial_{0} -
iA_{0}] + {\vec v}_{F}\cdot[i{\vec \nabla} + {\vec k}_{F} + {\vec
A}] \Bigr)\chi_{\sigma} \Bigr] \nn && +
\frac{2\sqrt{2d}a^{d-1}J_{K}}{c\tilde{g}}
\int_{0}^{c\beta}{dx_0}\sum_{k}
{\chi}_{k\sigma}^{\dagger}\tau^{3}_{\sigma\sigma'}\chi_{k+Q\sigma'}
. \eqa This action describes Fermi liquid physics of the
$\chi_{\sigma}$ fields under staggered magnetic fields. In the
strong Kondo coupling phase $J_{K}
>> J_{K}^{c}$ the Zeeman energy gap would be very large. This
leads the $\chi_{\sigma}$ band to be much flat, resulting in heavy
fermions $\chi_{\sigma}$. In this respect this phase is a heavy
fermion liquid for the Kondo resonances $\chi_{\sigma}$. {\it It
should be noted that the present heavy fermion liquid differs from
the usual one in that there is no analogous excitation with the
$\chi_{\sigma}$ field in the usual heavy fermion
liquid}\cite{Senthil_Kondo,Kim_Kondo}. A deeper analysis is
required to discriminate these two heavy fermion liquids. We
summarize the quantum phase transition from an antiferromagnetic
metal to a heavy fermion liquid in Table I.

Our present approach has some analogies with that of Ref.
\cite{Pepin_Kondo} in that bosonic spinons are utilized for
impurity spins and spinless fermions are introduced for Kondo
resonances. However, there are two important differences between
our approach and theirs. A critical field theory describing
deconfined quantum criticality necessarily has a new emergent
gauge structure in association with fractionalized fields. Our
critical field theory Eq. (8) allows U(1) gauge fields
guaranteeing the emergent U(1) gauge symmetry while the critical
theory in Ref. \cite{Pepin_Kondo} does not. Although use of the
U(1) gauge field is not a problem, it is not clear whether the
perturbative evaluations in Ref. \cite{Pepin_Kondo} keep the new
U(1) gauge symmetry in all steps. In addition, our spinless
fermions have two components while the ones in Ref.
\cite{Pepin_Kondo} have only one component corresponding to the
$\chi_{\uparrow}$ field in our representation for the spinless
fermions. Although we don't have any idea how the presence of two
components qualitatively alters the physics of one component at
present, we expect, at least, quantitative changes.

Lastly, we would like to point out that Fermi surface volume
continuously changes in the present description, consistent with
Ref. \cite{Pepin_Kondo}. In the present theory the staggered
Zeeman term controls the band of charged fermions. {\it Since the
staggered magnetic field continuously varies across the quantum
critical point, the fermion band smoothly changes, thus resulting
in continuous transformation of Fermi surface volume}.

In the present paper we claimed that the continuous quantum phase
transition in the Kondo lattice model can be realized by the
deconfined quantum criticality. Our critical field theory in Eq.
(6) not only explains non-Fermi liquid physics near the quantum
critical point but also recovers Fermi liquid physics away from
the quantum critical point.

We would like to thank Dr. Kim, Mun-Dae for helpful discussions
associated with his recent work in the Kondo lattice
model\cite{MunDae}.


\begin{thebibliography}{9}
\bibitem{Senthil_deconfinement} T. Senthil, A. Vishwanath, L. Balents, S. Sachdev, and M. P. A.
Fisher, Science {\bf 303}, 1490 (2004); T. Senthil, L. Balents, S.
Sachdev, A. Vishwanath, and M. P.A. Fisher, Phys. Rev. B {\bf 70},
144407 (2004).
\bibitem{Kim1} Ki-Seok Kim, Phys. Rev. B {\bf 72}, 035109
(2005).
\bibitem{Ichinose_deconfinement} D. Yoshioka, G. Arakawa, I. Ichinose, and
T. Matsui, Phys. Rev. B {\bf 70}, 174407 (2004); G. Arakawa, I.
Ichinose, T. Matsui, and K. Sakakibara, Phys. Rev. Lett. {\bf 94},
211601 (2005).
\bibitem{Review} P. A. Lee, N. Nagaosa, and X.-G. Wen,
cond-mat/0410445, and references therein.
\bibitem{Wen_QED3} W. Rantner and X.-G. Wen, cond-mat/0105540;
W. Rantner and X.-G. Wen, Phys. Rev. B {\bf 66}, 144501 (2002).
\bibitem{Hermele_QED3} M. Hermele, T. Senthil, M. P. A. Fisher,
P. A. Lee, N. Nagaosa, and X.-G. Wen, Phys. Rev. B {\bf 70},
214437 (2004).
\bibitem{Kleinert_QED3} F. S. Nogueira and H. Kleinert,
cond-mat/0501022.
\bibitem{Kim2} Ki-Seok Kim, Phys. Rev. B {\bf 70}, 140405(R)
(2004); Ki-Seok Kim, Phys. Rev. B {\bf 72}, 014406 (2005).
\bibitem{Kim3} Ki-Seok Kim, cond-mat/0502652.
\bibitem{Senthil_Kondo} T. Senthil, M. Vojta, and S. Sachdev,
Phys. Rev. B {\bf 69}, 035111 (2004); references therein; T.
Senthil, S. Sachdev, and M. Vojta, Phys. Rev. Lett. {\bf 90},
216403 (2003).
\bibitem{Coleman_Kondo}  P. Coleman, C. Pepin, Q. Si (Rice), and R.
Ramazashvili, Journal of Physics: Condensed Matter {\bf 13}, 723
(2001).
\bibitem{Pepin_Kondo} C. Pepin, Phys. Rev. Lett. {\bf 94},
066402 (2005).
\bibitem{Kim_Kondo} Ki-Seok Kim, Phys. Rev. B {\bf 71}, 205101
(2005).
\bibitem{Auerbach} A. Auerbach, Interacting Electrons and Quantum Magnetism, Springer (1998).
\bibitem{Nagaosa} N. Nagaosa, Quantum Field Theory in Strongly
Correlated Electronic Systems, Springer (1999).
\bibitem{Ladder} G. Sierra, J. Phys. A {\bf 29}, 3229 (1996).
\bibitem{Kim_1D} Ki-Seok Kim, cond-mat/0410231.
\bibitem{Chubukov} A. W. Sandvik, A. V. Chubukov, and S. Sachdev,
Phys. Rev. B {\bf 51}, 16483(R).
\bibitem{Kim_AF_ASL} Ki-Seok Kim, cond-mat/0504541.
\bibitem{Fisher_Z2} T. Senthil and M. P. A. Fisher,
Phys. Rev. B {\bf 62}, 7850 (2000).
\bibitem{Tesanovic} M. Franz and Z. Tesanovic,
Phys. Rev. Lett. {\bf 87}, 257003 (2001); This was pointed out by
prof. Tesanovic.
\bibitem{Fradkin} E. Fradkin, S. H. Shenker, Phys. Rev. D {\bf 19}, 3682
(1979).
\bibitem{NaLee} N. Nagaosa and P. A. Lee, Phys. Rev. B {\bf 61}, 9166
(2000).
\bibitem{Tsvelik_gauge} Alexei M. Tsvelik, Quantum Field Theory in
Condensed Matter Physics (Ch. 12), Cambridge University Press
(1995).
\bibitem{Nagaosa_gauge} P. A. Lee and N. Nagaosa, Phys. Rev. B {\bf 46},
5621 (1992).
\bibitem{Ioffe_gauge} L. B. Ioffe and G. Kotliar, Phys. Rev. B {\bf 42},
103408 (1990).
\bibitem{C_v1} T. Holstein, R. E. Norton, and P. Pincus, Phys. Rev.
B {\bf 8}, 2649 (1973).
\bibitem{C_v2} M. Yu. Reizer, Phys. Rev. B {\bf 40}, 11571 (1989).
\bibitem{Nagaosa_damping} N. Nagaosa, Phys. Rev. Lett. {\bf 71}, 4210
(1993).
\bibitem{Kleinert} H. Kleinert, F. S. Nogueira, and A. Sudbo, Phys. Rev. Lett. {\bf 88}, 232001 (2002);
H. Kleinert, F. S. Nogueira, and A. Sudbo, Nucl. Phys. B {\bf
666}, 361 (2003).
\bibitem{MunDae} M. D. Kim, C. K. Kim, and J. Hong,
Phys. Rev. B {\bf 68}, 174424 (2003).
\end{thebibliography}
\end{document}